\documentclass[journal]{IEEEtran}

\usepackage{cite,graphicx,amsmath,amssymb}
\usepackage{subfigure}
\usepackage{citesort}
\usepackage{fancyhdr}
\usepackage{mdwmath}
\usepackage{mdwtab}
\usepackage{balance}
\usepackage{xcolor}
\usepackage{bm}
\usepackage{amsthm}
\usepackage{threeparttable}
\usepackage{algorithm}
\usepackage{algorithmic}
\usepackage{multirow}
\usepackage{flafter}
\usepackage{makecell}
\usepackage{diagbox}

\providecommand{\url}[1]{#1}

\begin{document}
\title{Semantic Communications in Multi-user Wireless Networks}
\author{
Xidong~Mu and Yuanwei~Liu

\thanks{Xidong Mu and Yuanwei Liu are with the School of Electronic Engineering and Computer Science, Queen Mary University of London, London E1 4NS, UK, (email: \{xidong.mu, yuanwei.liu\}@qmul.ac.uk).}
}


\maketitle
\begin{abstract}
This article investigates the exploitation of semantic communications in multi-user networks. We propose a novel heterogeneous semantic and bit multi-user framework for providing flawless, customized, and intelligent information transmission. We discuss both orthogonal multiple access (OMA) and non-orthogonal multiple access (NOMA) for the proposed heterogeneous framework, with an emphasis on investigating the attractive interplay between semantic communications and NOMA, namely \emph{NOMA enabled semantic communications} and \emph{semantic communications enhanced NOMA}. 1) For NOMA enabled semantic communications, we propose a semi-NOMA scheme for efficiently facilitating the heterogeneous semantic and bit multi-user communication, which unifies conventional NOMA and OMA schemes. The fundamental performance limit, namely semantic-versus-bit rate region, is characterized, which shows the superiority of the proposed semi-NOMA. 2) For semantic communications enhanced NOMA, we propose an opportunistic semantic and bit communication approach to alleviate the early-late rate disparity issue in NOMA. Numerical case studies demonstrate that significant performance gain can be achieved for NOMA by employing semantic communications than bit communications. Finally, several open research directions are highlighted.
\end{abstract}

\section{Introduction}

Following the pioneering contributions of the Shannon classical information theory~\cite{Shannon}, wireless communications have significantly developed from the first generation (1G) to the fifth generation (5G) only in a few decades. For approaching the fundamental Shannon rate limit, diverse efficient technologies have been developed, such as massive multiple-input multiple-output (MIMO), mmWave/THz communication, and turbo codes~\cite{6736746}. Nevertheless, the development of wireless communication is far from the end and extensive research efforts have been devoted to developing the next generation (6G) wireless communication system~\cite{Tong,6G}. This is because, on the one hand, the requirements of communication performance are still explosively growing (e.g., Tbps-level data rates, 0.1 millisecond-level latency, 99.99999\% reliability, etc.). On the other hand, the communication services have to be provided for both conventional human-type users and new machine-type users as well as Internet-of-Things (IoT) users. To achieve these goals, the effectiveness of relentlessly increasing of the size of transceivers and the frequencies of operating carriers is questionable given the extremely high hardware cost and energy consumption. Moreover, one important observation is that current wireless communication designs still focus on the technical-level problem~\cite{Shannon} using the Shannon classical information theory, i.e., \emph{How accurately can the symbols of communication be transmitted?}. There is a paucity of investigations on the other two semantic-level and effectiveness-level problems~\cite{Shannon}, i.e., \emph{How precisely do the transmitted symbols convey the desired meaning?} and \emph{How effectively does the received meaning affect conduct in the desired way?}, which are more related to the ultimate goal of communications. 
\begin{figure*}[htb]
  \centering
  \includegraphics[width= 6in]{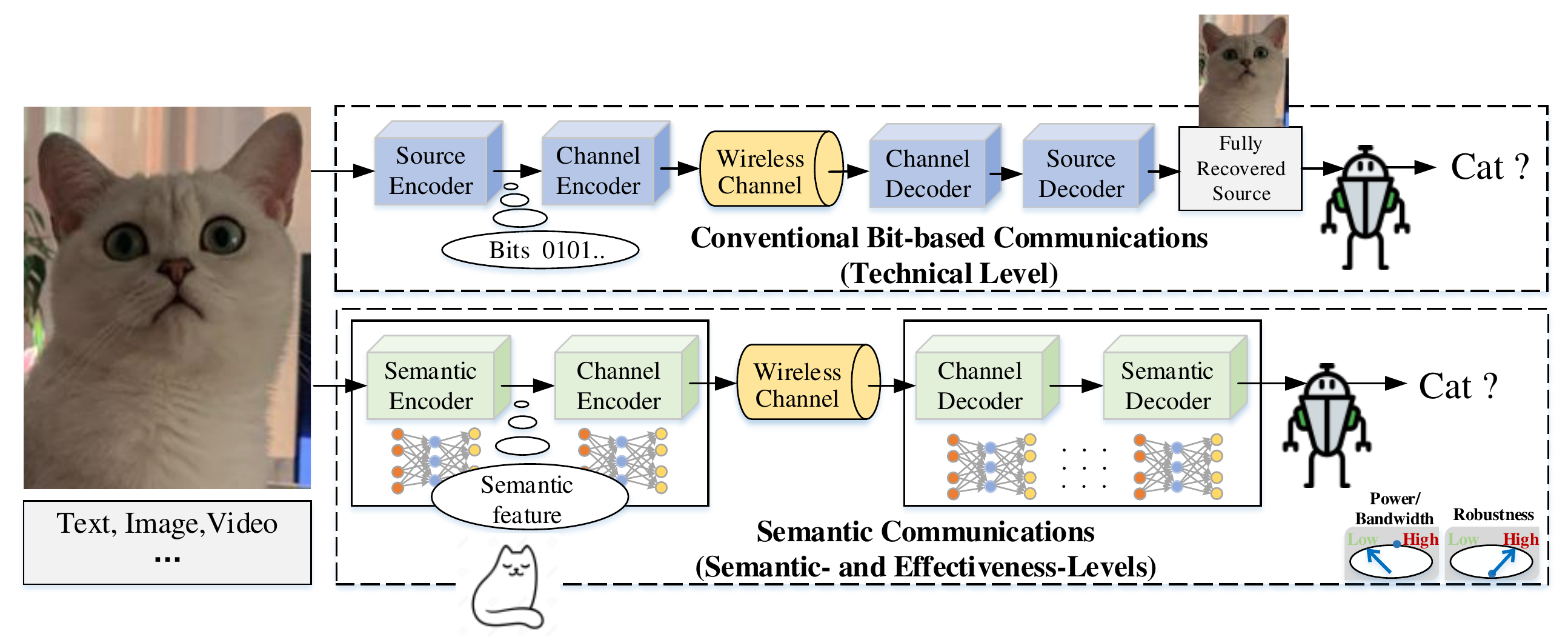}\\
  \caption{Illustration of conventional bit-based communications and new semantic communications.}\label{figure1}
\end{figure*}

In response to the semantic- and effectiveness-level communication problems, semantic communications have recently attracted significant research attentions from both industry and academia~\cite{Qin,Gunduz}. In semantic communications, only the key information that contains the specific meaning/actions/goals relevant to the destinations needs to be transmitted. This is different from conventional bit-based communications, the entire sources have to be transmitted~\cite{Qin}. Fig. \ref{figure1} illustrates the difference between conventional bit-based communications and semantic communications in an end-to-end scenario. Here, we take the image recognition as an example. The transmitter in conventional bit-based communications sends the bit sequences converted from the entire image, which allows the receiver to fully recover the image for recognizing the cat. By contrast, the transmitter in semantic communications only needs to transmit the key features extracted by the semantic encoder. Upon receiving the semantic symbols, the semantic decoder enables the receiver to directly recognize that the received information is a cat. Compared to the conventional bit-based communications, the abundant information (e.g., image background and colors) for the image recognition is safely dropped in semantic communications, thus focusing on the semantic- and effectiveness-level design~\cite{Qin}. Given the significant breakthrough of artificial intelligence in semantic information extraction, many efficient approaches have been developed for facilitating semantic communications, such as text/speech/image/video transmission~\cite{DeepSC,9450827,Kang}, image classification, machine translation, and visual question answering~\cite{Xie}. The salient advantages of semantic communications are listed as follows:   
\begin{itemize}
  \item \textbf{Sustainability}: As only the necessary information needs to be transmitted, the power consumption and bandwidth required in semantic communications can be greatly reduced, thus leading to a sustainable and green communication paradigm. 
  \item \textbf{Reliability}: Semantic communications exhibit robustness in undesired wireless environment, e.g., in the low signal-to-noise ratio (SNR) regime, thus ensuring the reliability of communication systems.
  \item \textbf{Intelligence}: The information transmitted in semantic communications depends on the specific tasks, thus achieving intelligent information exchange. 
\end{itemize}  

Motivated by these benefits, we focus our attentions on exploiting semantic communications in multi-user networks for building up flawless, customized, and intelligent information flow among users. The integration of semantic communications into multi-user networks leads to highly heterogeneous features. On the one hand, from the network perspective, both new semantic-based users (S-users) and existing bit-based users (B-users) have to be accommodated. On the other hand, from the user perspective, both semantic and bit communication approaches can be employed for information transmission. This calls for efficient multiple access and resource management for supporting the heterogeneous semantic and bit multi-user communications, which provides the main motivation of this article.

The rest of this article is organized as follows. A novel heterogeneous semantic and bit multi-user framework is proposed employing orthogonal multiple access (OMA) or non-orthogonal multiple access (NOMA), followed by discussing the promising interplay between semantic communications and NOMA. 1) In the context of ``NOMA enabled semantic communications'', a semi-NOMA enabled heterogeneous semantic and bit multi-user communication design is proposed. 2) In the context of ``semantic communications enhanced NOMA'', a novel opportunistic semantic and bit communication approach is proposed for alleviating the early-late rate disparity issue in NOMA. For each proposed design, numerical examples are provided for verifying the effectiveness. Finally, conclusions are drawn and several future directions are discussed.
\begin{figure}[htb]
  \centering
  \includegraphics[width= 3in]{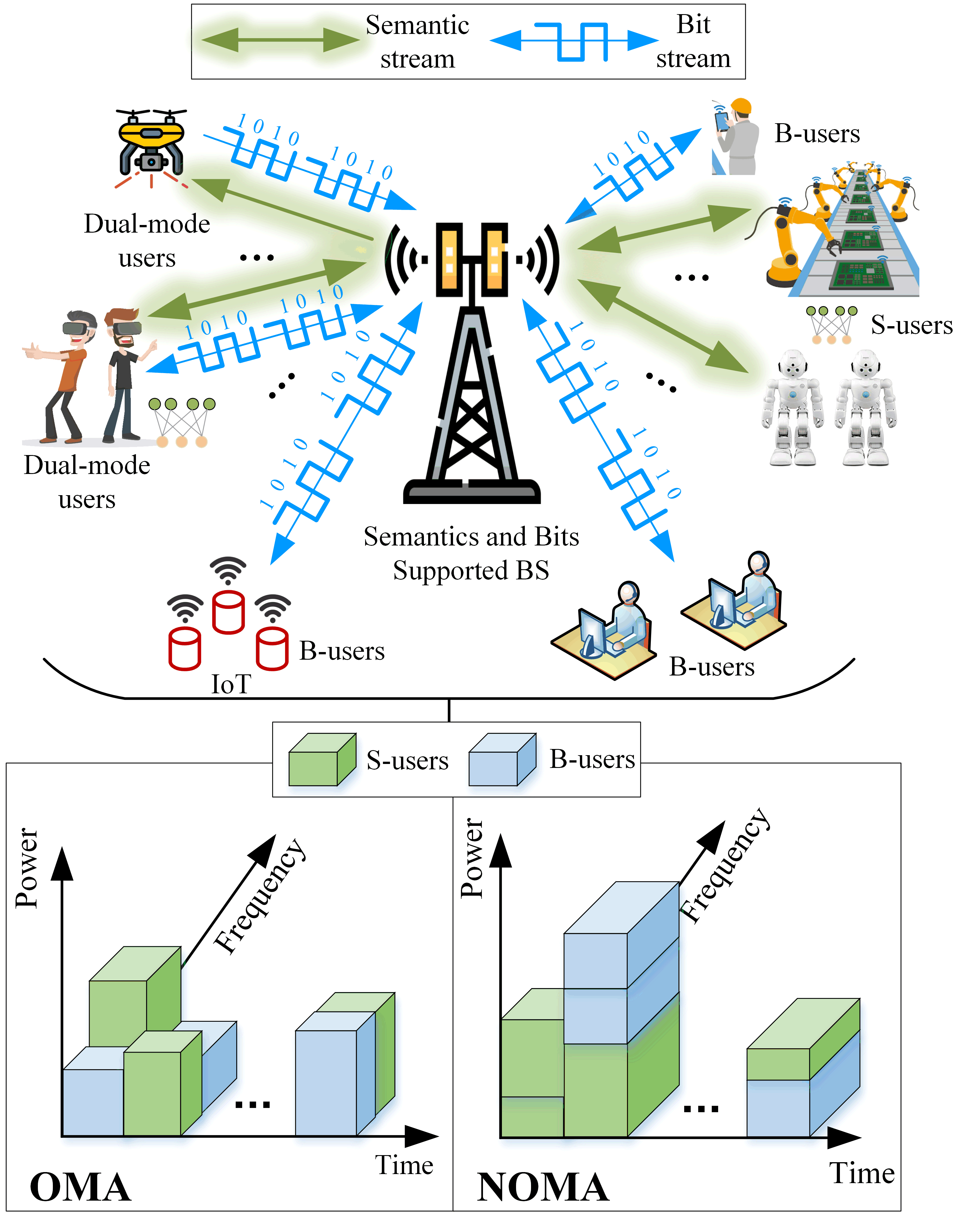}\\
  \caption{The proposed heterogeneous semantic and bit multi-user framework employing OMA and NOMA.}\label{figure2}
\end{figure}
\section{Heterogeneous Semantic and Bit Multi-user Frameworks}
Fig. \ref{figure2} illustrates the proposed heterogeneous semantic and bit multi-user framework, which consists of a semantics and bits supported base station (BS) to serve both the conventional B-users and the new emerging S-users for facilitating different applications. As illustrated, S-users can be robotic users in smart factories for task executions, while B-users are conventional IoT devices for data collections or laptops for file transmission. Moreover, users can also work in a dual semantic and bit communication mode. For example, the BS control an unmanned aerial vehicle (UAV) for video monitoring via semantic communications and the UAV sends the video back to the BS via bit communications considering the high-quality air-to-ground link. Moreover, for virtual reality (VR)/augmented reality (AR) users in immersive games, semantic and bit communications would coexist for supporting different information flows and realizing multimodal experience (e.g., video, control, haptic perception, etc.). Therefore, compared to current bit-only networks, the proposed heterogeneous semantic and bit multi-user framework enables the information transmission to be realized in the most efficient manner possible in terms of different applications.

\subsection{Multiple Access for the Proposed Heterogeneous Semantic and Bit Multi-user Framework}
For facilitating the proposed heterogeneous semantic and bit multi-user framework, multiple access is one of the most fundamental enablers given the limited wireless resources. Here, we broadly classify the multiple access schemes into two categories, namely OMA and NOMA. 
\begin{itemize}
\item \textbf{OMA}: On the one hand, OMA can be employed in the proposed heterogeneous multi-user framework by allocating S- and B-users with different orthogonal time/frequency resource blocks, as illustrated in Fig. \ref{figure2}. Like conventional bit-only multi-user communications, the advantages of OMA are that there is no interference between each user and easy for implementation. Therefore, no redesigns is required in the individual semantic and bit communications. However, OMA usually has a low resource efficiency and cannot support massive connectivity. This shortcoming of OMA will become more significant when involving new S-users in the proposed heterogeneous multi-user framework.  
\item  \textbf{NOMA}: On the other hand, as illustrated in Fig. \ref{figure2}, NOMA can accommodate S- and B-users over the same time/frequency resource blocks and increase the connectivity performance for the proposed heterogeneous multi-user framework. Therefore, NOMA is more promising to be employed than OMA. However, note that the multi-user interference mitigation is the core task in NOMA~\cite{NOMA}. How to efficiently use NOMA and design the corresponding multi-user interference mitigation in the heterogeneous framework are challenging tasks. 
\end{itemize} 
\subsection{Interplay between Semantic Communications and NOMA}
Against the above discussion, in this paper, we focus our attentions on the employment of NOMA in the proposed heterogeneous semantic and bit multi-user framework. In particular, the attractive interplay between semantic communications and NOMA can be summarized as follows.  
\begin{itemize}
  \item \textbf{NOMA enabled Semantic Communications}: On the one hand, NOMA provides efficient and flexible resource allocation for incorporating semantic communications and improving the resource efficiency and connectivity of the proposed heterogeneous multi-user framework. This constitutes the most foundation for facilitating diversified applications in future wireless networks, such as VR, AR, and Metaverse. The details of NOMA enabled heterogeneous semantic and bit multi-user communications will be discussed in Section III.
  \item \textbf{Semantic Communications enhanced NOMA}: On the other hand, the new information transmission paradigm provided by semantic communications can also improve the performance of NOMA. As discussed above, semantic communications are more robust and sustainable than conventional bit communications. Applying semantic communications in NOMA can enhance the performance of users having a poor channel conditions and/or reduce the required transmit power at users, i.e., reducing the multi-user interference for NOMA. This enables us to employ semantic communications for improving NOMA performance, the details of which will be discussed in Section IV. 
\end{itemize} 

\section{NOMA enabled Heterogeneous Semantic and Bit Multi-user Communications}

\begin{figure*}[htb]
  \centering
  \includegraphics[width= 6in]{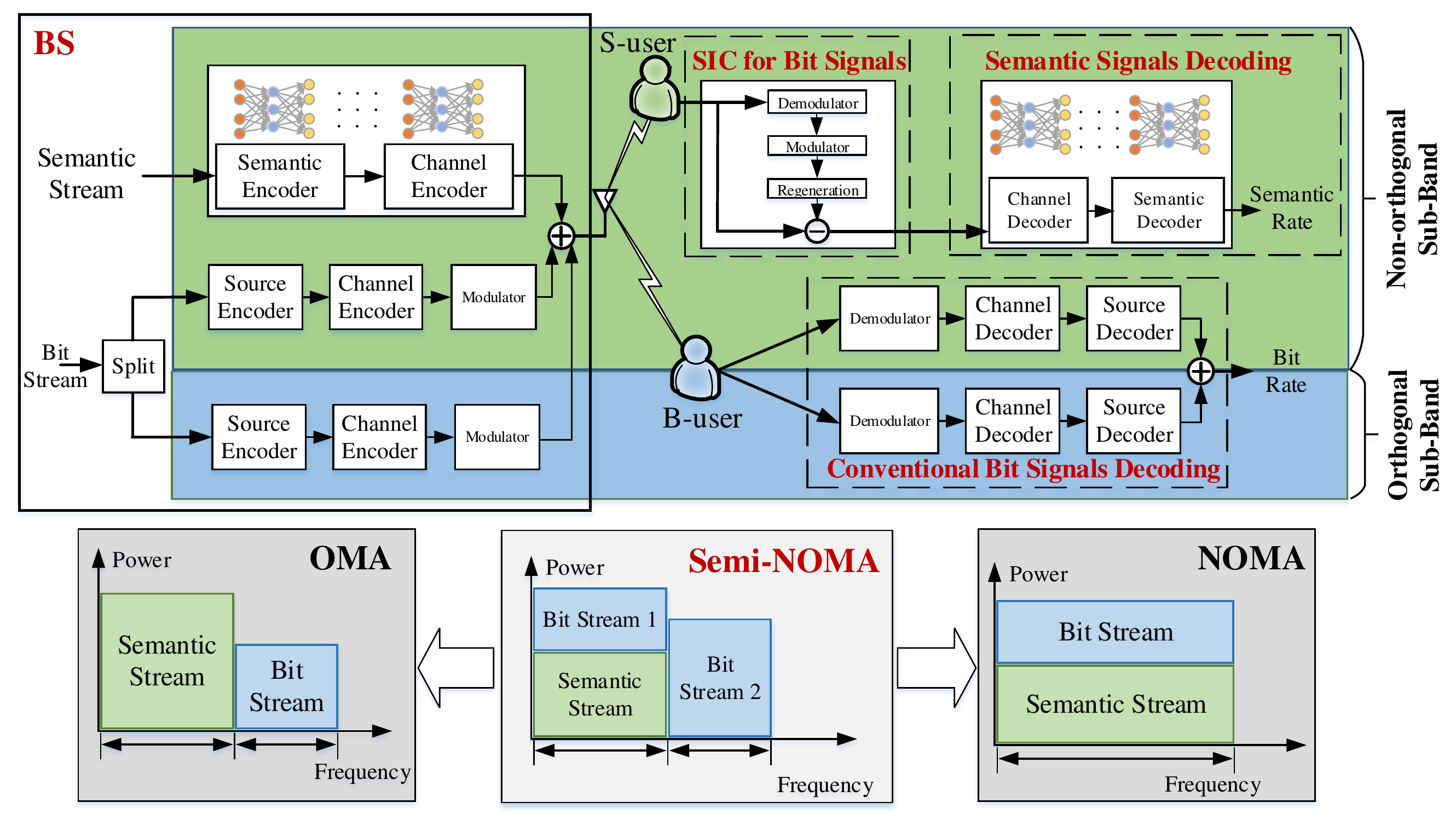}\\
  \caption{Semi-NOMA for heterogeneous semantic and bit communications.}\label{figure3}
\end{figure*}
In this section, we focus on ``NOMA enabled semantic communications''. We will first introduce the new performance metric, namely semantic rate, for characterizing the performance of semantic communications. Then, we propose a novel semi-NOMA scheme for realizing the heterogeneous semantic and bit multi-user communications, which is capable of achieving the best semantic-versus-bit rate region.
\subsection{Semantic Rate: A New Performance Metric}
Conventional bit communications focus on the generalized performance metrics, such as symbol-error rate (SER) and bit-error-rate (BER), regardless of the types of original sources. The classic Shannon capacity characterizes the performance upper bound of bit communications by ignoring the specific meaning contained in the bit sequences. However, as semantic communications focus on the correctness of information transmission in the semantic domain, new performance metrics have to be developed for characterizing its performance. Depending on the different types of sources (e.g., text, speech, image) and different tasks/goals to be accomplished, the corresponding performance metrics for semantic communications are in general different. This also imposes challenges in multiple access design and resource management when incorporating semantic communications into multi-user networks. To this end, based on one practical DL-enabled semantic communication (DeepSC) tool proposed in~\cite{DeepSC} for text transmission, the authors of \cite{Yan} proposed a new performance metric, namely semantic rate, to quantify the efficiency of semantic communications. In particular, the achieved semantic rate depends on the employed bandwidth and the achieved sentence semantic similarity, which is a function of the received signal-to-noise ratio (SNR) and the specific neural network structure used in the DeepSC. However, in \cite{Yan}, there is no closed-form expressions provided for the sentence semantic similarity function, the value of which can only be obtained via experiments on DeepSC. To address this difficulty, we further proposed to employ the data regression method, where the `S-shape' sentence semantic similarity function was approximated by a generalized logistic function with a high accuracy \cite{Mu_JSAC1}. This facilitates further theoretical studies and communication designs on semantic communications. It is also worth noting that the development of performance metrics for semantic communications is still in an initial stage, which requires further research efforts for different types of semantic communications. 

\subsection{Semi-NOMA: A Unified Multiple Access Scheme}
As shown in Fig. \ref{figure3}, we focus on the downlink case, where a BS aims to transmit the semantic stream and the bit stream to one S-user and B-user over the given wireless resource. Based on the introduced semantic rate, we investigate how to efficiently support this heterogeneous multi-user communication with the employment of NOMA. Previously, the superiority of NOMA over OMA has been extensively demonstrated in bit-only multi-user communications. However, this conclusion may not hold in the new heterogeneous scenario and proper redesigns are required for exploiting the NOMA gain. Recalling the fact that for downlink NOMA in conventional bit communications, the successive interference cancelation (SIC) decoding order among users for mitigating the interference caused by spectrum sharing is determined by the order of users' channel conditions \cite{NOMA}. Users having higher channel conditions will first decode the signal of users having lower channel conditions and remove them from the superimposed signal. However, this `channel condition-based SIC ordering' cannot be applied in the new heterogeneous multi-user communication. This is because for semantic communications, the neural networks employed at the BS and the S-user should be jointly trained in advance~\cite{DeepSC}, thus ensuring the successful implementation. As a result, the B-user in the heterogeneous transmission is almost impossible to decode the semantic signal from the superimposed semantic and bit signal and thus the SIC cannot be carried out. Nevertheless, as the bit communications do not require the prior training, S-user is still capable of decoding the bit signal from the superimposed semantic and bit signal. This leads to a `bits-to-semantics SIC ordering' when applying NOMA in the heterogeneous multi-user communication. However, despite the S-user can decode its intended semantic signal in an interference-free manner with the aid of SIC, the drawback is that the B-user will always suffer from the interference from the S-user. Therefore, if NOMA is directly applied in the heterogeneous multi-user communication (termed as pure NOMA), the performance of the B-user might cannot be well guaranteed.  

To guarantee the performance of both S- and B-users, we propose a novel semi-NOMA scheme for realizing the heterogeneous multi-user communication~\cite{Mu_JSAC1}. As shown in Fig. \ref{figure3}, the total bandwidth is divided into one non-orthogonal sub-band and one orthogonal sub-band. Based on this bandwidth allocation, the operating mechanism of each node in the proposed semi-NOMA scheme is summarized as follows.
\begin{itemize}
  \item \textbf{BS}: As illustrated, the bit stream for the B-user should be split into two sub-streams at the BS, where one bit sub-stream is superimposed with the semantic stream and transmitted over the non-orthogonal sub-band while the other bit sub-stream is transmitted alone over the orthogonal sub-band.   
  \item \textbf{S-user}: Upon receiving the superimposed from the non-orthogonal sub-band, the S-user decodes the superimposed signal following the aforementioned bits-to-semantics SIC ordering. As a result, the semantic signal is still decoded in an interference-free manner in semi-NOMA.
  \item \textbf{B-user}: Accordingly, the B-user decodes the two bit sub-streams received from the non-orthogonal and orthogonal sub-bands and combines the decoded results. It can be observed that compared to pure NOMA, partial information transmission of B-user in semi-NOMA can also be achieved in an interference-free manner. 
\end{itemize} 

The resource allocation among S- and B-users in the proposed semi-NOMA is illustrated in the bottom of Fig \ref{figure3}. It can be observed that by adjusting the bandwidth and power allocation, the proposed semi-NOMA can be simplified into OMA and NOMA. Notably, semi-NOMA also provides more transmission options, which cannot be achieved by OMA and NOMA. Therefore, the salient advantage of semi-NOMA is that it provides a unified multiple access scheme for facilitating the heterogeneous semantic and bit multi-user communication. 

\subsection{Fundamental Limit: Semantic-versus-Bit Rate Region}
\begin{figure}[htb]
  \centering
  \includegraphics[width= 3.5in]{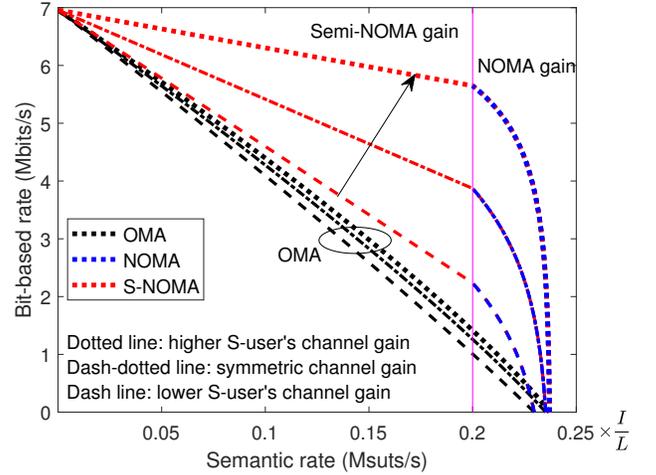}\\
  \caption{Illustration of the achieved semantic-versus-bit rate region achieved by different multiple access schemes. The system parameter settings can be found in \cite[Section VI]{Mu_JSAC1}.}\label{figure4}
\end{figure}

We continue to investigate the fundamental performance limit of the considered heterogeneous semantic and bit multi-user communication, namely the semantic-versus-bit rate region, which characterizes all the semantic and bit rate-pairs achieved by S- and B-users for any given transmit power and bandwidth~\cite{Mu_JSAC1}. Fig. \ref{figure4} compares the semantic-versus-bit rate region achieved by the proposed semi-NOMA and the conventional OMA and NOMA. In the provided numerical results, we fix the channel condition of the B-user and vary the channel condition of the S-user~\cite{Mu_JSAC1}. Therefore, compared to the channel condition of the B-user, we have three cases: (1) a higher channel condition at the S-user, (2) a symmetric channel condition, and (3) a lower channel condition at the S-user. As can be observed from Fig. \ref{figure4}, in all three cases, semi-NOMA achieves the largest semantic-versus-bit rate region, which strictly contains those achieved by OMA and NOMA. This is indeed expected since semi-NOMA unifies OMA and NOMA and provides enhanced degrees-of-freedom for resource allocation in the heterogeneous multi-user communication. It can be also observed that due to the full resource sharing, NOMA is only capable of achieving a reduced semantic-versus-bit rate region, where the maximum bit rate is degraded by the interference from the S-user. Moreover, an interesting result is that even for the case of symmetric channel condition, OMA is still strictly suboptimal. This is fundamentally different from the result in the conventional bit-only multi-user communication, where OMA achieves the same performance as NOMA when users have the same channel conditions. This reveal the importance of using NOMA-based schemes when including semantic communications in multi-user network. By comparing the achieved semantic-versus-bit rate region in the three cases, we can find that it is preferable to pair an S-user having a higher channel condition with a B-user having a lower channel condition in semi-NOMA, which leads to a larger semantic-versus-bit rate region. This provides a useful guidance for implementing semi-NOMA. The above results confirm the effectiveness of the proposed semi-NOMA scheme in the considered heterogeneous semantic and bit multi-user communication.

\section{Semantic Communications Enhanced NOMA}
In this section, we continue to discuss the employment of semantic communications for improving the performance of NOMA, i.e., ``semantic communications enhanced NOMA''. We first introduce the ``early-late'' rate disparity issue among NOMA users. Then, we explore semantic communications to alleviate this issue and propose a novel opportunistic semantic and bit communication approach for NOMA users.
\subsection{Early-Late Rate Disparity Issue in NOMA}
Due to the resource sharing nature of NOMA, the mitigation of multi-user interference is the core task for further exploiting the limited wireless resources and supporting massive connectivity. To this end, NOMA relies on the employment of SIC to deal with the multi-user interference \cite{NOMA}. For implementing the SIC, users should be priorly ordered in terms of specific criteria, such as the order of users' channel conditions and the order of users' quality-of-service (QoS) requirements~\cite{SIC}. As a result, for any given SIC decoding order, users who are assigned at earlier decoding orders will always suffer from more multi-user interference than those who are assigned at later decoding orders, which yields the early-late rate disparity issue among NOMA users. In other words, the communication performance achieved by each NOMA user significantly varies in terms of their SIC decoding orders. 

\begin{figure*}[htb]
  \centering
  \includegraphics[width= 6in]{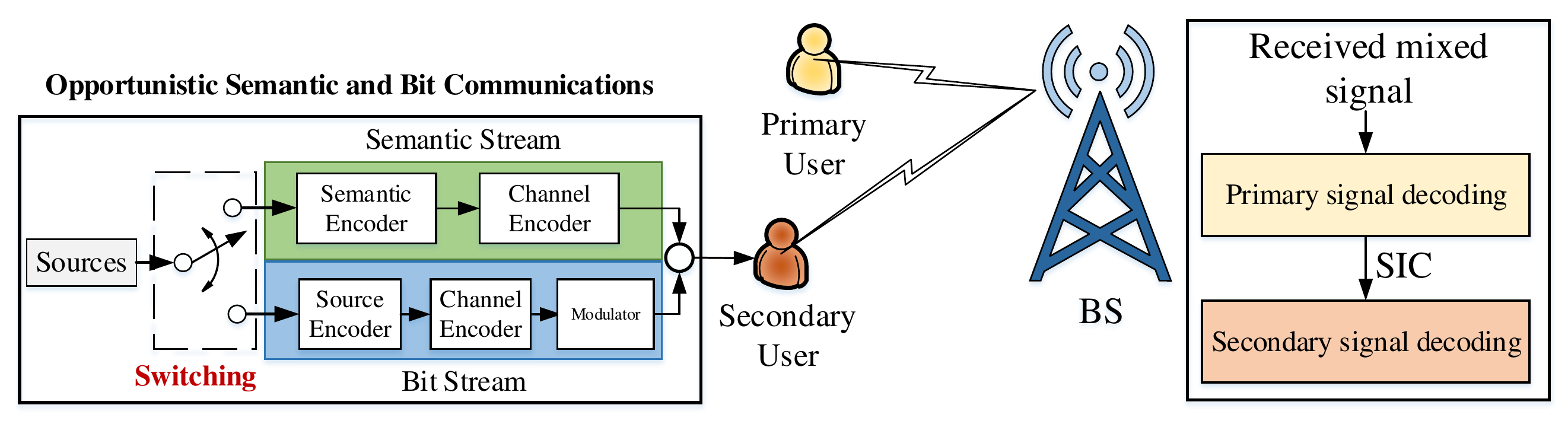}\\
  \caption{The proposed opportunistic semantic and bit communication approach for NOMA.}\label{figure5}
\end{figure*}

To further demonstrate the limitation caused by the early-late rate disparity issue, as shown in Fig. \ref{figure5}, we consider a fundamental two-user uplink NOMA system, which includes one primary user and one secondary user. At the BS, the signal of the primary user is firstly decoded and then removed using SIC before the signal of the secondary user is decoded. Therefore, on the one hand, it is discouraged for the primary user to participate in NOMA since its performance will be degraded by the interference from the secondary user. On the other hand, although the secondary user's signal can be decoded in an interference-free manner, its maximum achieved performance is also conditioned on the communication requirement of the primary user. For example, if the primary user has a stringent communication requirement, the signal strength of the secondary user should be strictly capped for controlling the interference. Owing to the early-late rate disparity issue, it is impossible for the two NOMA users to simultaneously achieve high communication performance. As a result, previous works usually paired users with different QoS requirements together when using NOMA~\cite{NOMA,SIC}. Nevertheless, the early-late rate disparity issue in NOMA has not been solved.

\subsection{An Opportunistic Semantic and Bit Communication Approach for NOMA}
Current research contributions have revealed that for achieving the same communication performance, semantic communications consumes less transmit power than conventional bit communications~\cite{Qin}. It implies that semantic communications would cause less multi-user interference without degrading the communication performance. Motivated by this observation, as shown in Fig. \ref{figure5}, we propose an opportunistic semantic and bit communication approach for NOMA users. By switching between the two modes, the original sources at the user are converted into either the semantic stream or the bit stream. The proposed opportunistic approach enables the user to employ the most suitable communication strategy for participating in NOMA with other users. Let us still take the aforementioned two-user as an example. Intuitively, if the primary user has a stringent communication requirement and/or the secondary user experiences a poor channel condition, the employment of semantic communications is desired. Otherwise, the employment of bit communications would be desired. 

\subsection{Numerical Case Studies}
\begin{figure}[htb]
  \centering
  \includegraphics[width= 3.5in]{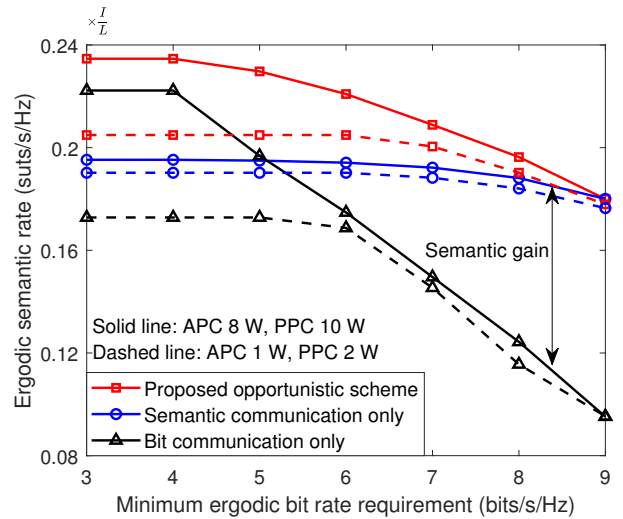}\\
  \caption{Illustration of the ergodic (equivalent) semantic rate achieved by the secondary user versus the ergodic bit rate required by the primary user in different schemes. The system parameter settings can be found in \cite[Section V]{Mu_JSAC2}.}\label{figure6}
\end{figure}

We continue to provide numerical examples to show the performance gain of employing semantic communications for NOMA and the effectiveness of the proposed opportunistic communication approach. The simulation results are obtained by considering a primary user using bit communications and a secondary user using the proposed opportunistic semantic and bit communication approach~\cite{Mu_JSAC2}. We study the NOMA performance over fading channels, where the secondary user has to choose the desired strategy in each fading state. For comparison, two baselines are also considered, where the secondary user merely uses semantic communications or bit communications. Moreover, if the secondary user employs bit communications, the corresponding achieved bit rate is converted into the equivalent semantic rate for a fair comparison. Fig. \ref{figure6} shows the ergodic (equivalent) semantic rate achieved by the secondary user over fading channels versus the ergodic bit rate required by the primary user, subject to the average power constraint (APC) and the peak power constraint (PPC) of the secondary user. It can be observed that by reaping the benefits of both semantic and bit communications, the proposed opportunistic scheme achieves the best performance for NOMA. It allows the secondary user not only to efficiently control the interference imposed to the primary user but also to enhance it own communication performance. We can also observe that bit communications are preferable to be employed when the primary user has a low communication requirement and the secondary user has a sufficiently large power budget. Moreover, when the primary user has a stringent communication requirement and/or the secondary user has limited transmit power, the employment of semantic communications can achieve a significant performance gain for NOMA than bit communications. The provided results confirm the effectiveness of employing semantic communications for enhancing the performance of NOMA.

\section{Concluding Remarks and Future Research}
In this article, a novel heterogeneous semantic and bit multi-user framework was proposed by incorporating newly emerging semantic communications. Compared to OMA, it was reveled that the interplay between NOMA and semantic communications has great potentials for supporting the proposed heterogeneous multi-user framework, namely NOMA enabled semantic communications and semantic communications enhanced NOMA. For NOMA enabled semantic communications, a semi-NOMA enabled heterogeneous semantic and bit communication scheme was proposed, which unifies conventional OMA and NOMA as special cases and provides more flexible transmission options. It was also numerically shown that the proposed semi-NOMA scheme can achieve the largest semantic-versus-bit rate region. For semantic communications enhanced NOMA, an opportunistic semantic and bit communication approach was conceived for alleviating the early-late rate disparity issue in NOMA. Numerical studies showed that the employment of semantic communications enables the NOMA users to better balance between its own achieved performance and the imposed interference, i.e., further improving the NOMA performance. Note that there are still numerous open problems for employing semantic communications and NOMA in the proposed heterogeneous multi-user framework, some of which are listed as follows.    
\begin{itemize}
  \item \textbf{Machine Learning Empowered Resource Allocation}: Given the large scales and massive connectivity of future multi-user networks, the resulting resource allocation problems in the proposed heterogeneous semantic and bit multi-user framework (e.g., NOMA user scheduling and communication strategy switching) will become extremely complicated and challenging. Instead of conventional convex optimization methods, machine learning methods are envisioned as powerful tools to address these issues. Therefore, the development of efficient machine learning empowered resource allocation algorithms constitute a promising research direction.
  \item \textbf{Security Provisioning}: The security issues are of great significance in the proposed heterogeneous multi-user framework, especially when considering the resource sharing nature of NOMA. For security protection in semantic communications, on the one hand, conventional physical layer security (PLS) methods can be employed to combat eavesdropping. This, however, requires the redefinition of the PLS measurements for the new beyond-bits semantic communications. On the other hand, the secrecy performance of semantic communications under adversarial attacks remains largely unknown, which requires further research.
  \item \textbf{Proof-of-Concept Design and Verification}: To validate the obtained theoretical performance gains, real implementations of the proposed heterogeneous multi-user framework are necessary. However, for the proof-of-concept design, several practical tasks should be focused on, including the mitigation of error propagation in SIC when employing NOMA, the design of modulation and detection approaches in transceivers, and the effective training of neural networks for different S-users. These problems merit further investigation.
\end{itemize}


\begin{thebibliography}{1}

\bibitem{Shannon}
W.~{Weaver} and C.~E. {Shannon}, \emph{The Mathematical Theory of Communication}.\hskip 1em plus 0.5em minus 0.4em\relax University of Illinois Press, 1949.

\bibitem{6736746}
F.~Boccardi, \emph{et al.}, ``Five disruptive technology directions for {5G},'' \emph{{IEEE} Commun. Mag.}, vol.~52, no.~2, pp. 74--80, 2014.

\bibitem{Tong}
W. Tong and G. Y. Li, ``Nine Challenges in Artificial Intelligence and Wireless Communications for 6{G},'' \emph{{IEEE} Wireless Commun.}, vol. 29, no. 4, pp. 140-145, 2022.

\bibitem{6G}
G.~Liu, \emph{et al.}, ``Vision, requirements and network architecture of 6{G} mobile network beyond 2030,'' \emph{China Communications}, vol. 17, no. 9, pp. 92-104, 2020.

\bibitem{Qin}
Z.~Qin, \emph{et al.}, ``Semantic communications: Principles and challenges,'' \emph{arXiv:2201.01389}, 2022.

\bibitem{Gunduz}
D.~{Gunduz}, \emph{et al.}, ``Beyond transmitting bits: Context, semantics and task-oriented communications,'' \emph{arXiv:2207.09353}, 2022.

\bibitem{DeepSC}
H.~Xie, \emph{et al.}, ``Deep learning enabled semantic communication systems,'' \emph{{IEEE} Trans. Signal Process.}, vol.~69, pp. 2663--2675, 2021.

\bibitem{9450827}
Z.~Weng and Z.~Qin, ``Semantic communication systems for speech transmission,'' \emph{{IEEE} J. Sel. Areas Commun.}, vol.~39, no.~8, pp. 2434--2444, 2021.

\bibitem{Kang}
J.~Kang, \emph{et al.}, ``Personalized Saliency in Task-Oriented Semantic Communications: Image Transmission and Performance Analysis,'' \emph{{IEEE} J. Sel. Areas Commun.}, accept to appear, \emph{arXiv:2209.12274}, 2022.

\bibitem{Xie}
H.~Xie, \emph{et al.}, ``Task-Oriented Multi-User Semantic Communications,'' \emph{IEEE J. Sel. Areas Commun.}, vol. 40, no. 9, pp. 2584-2597, 2022.

\bibitem{NOMA}
Y. Liu, \emph{et al.}, ``Nonorthogonal multiple access for 5{G} and beyond,'' \emph{Proc. {IEEE}}, vol. 105, no.~12, pp. 2347--2381, 2017.

\bibitem{Yan}
L.~Yan, \emph{et al.}, ``Resource allocation for text semantic communications,'' \emph{{IEEE} Wireless Commun. Lett.}, vol.~11, no.~7, pp. 1394--1398, 2022.

\bibitem{Mu_JSAC1}
X.~Mu, \emph{et al.}, ``Heterogeneous Semantic and Bit Communications: A Semi-{NOMA} Scheme,'' \emph{IEEE J. Sel. Areas Commun.}, accept to appear, \emph{arXiv:2205.02620}, 2022.

\bibitem{SIC}
Z. Ding, \emph{et al.}, ``Unveiling the importance of {SIC} in {NOMA} systems—part 1: State of the art and recent findings,'' \emph{{IEEE} Commun. Lett.}, vol.~24, no.~11, pp. 2373--2377, 2020.

\bibitem{Mu_JSAC2}
X.~Mu and Y. Liu, ``Exploiting Semantic Communication for Non-Orthogonal Multiple Access,''  \emph{arXiv:2209.06006}, 2022.


\end{thebibliography}
\end{document}